# Emergent spatial correlations in stochastically evolving populations

Marek Grabowski and R.E. Camley

*Department of Physics, University of Colorado, Colorado Springs, Colorado 80933-7150*

We study the spatial pattern formation and emerging long range correlations in a model of three species coevolving in space and time according to stochastic contact rules. Analytical results for the pair correlation functions, based on a truncation approximation and supported by computer simulations, reveal emergent strategies of survival for minority agents based on *selection of patterns*. Minority agents exhibit *defensive clustering* **and** *cooperative behavior* **close to phase transitions.**

The dynamics and spatial pattern formation of interacting species have been recently studied in various contexts[1,2] using both deterministic and stochastic modelling techniques. Particular emphasis in these studies was placed on temporal evolution of global quantities[3] such as population densities. In contrast, much less is known about the evolution of multi-species systems in space, especially when the local rules are probabilistic rather than deterministic (cellular automata[4]). There has been a growing recognition[5], however, that the environment has a spatial dimension, since individual population members rarely mix homogeneously over the territory available to them but develop instead within separate sub-regions. It is in this context that explicitly spatial stochastic versions of the classical models[6,7] have received renewed attention[3,8,9].

Specifically, spatial patterns in chemical reaction systems were investigated[8,9] in the mean-field and pair approximations, and contrasted with stochastic simulation results. Active stationary states of oscillating populations[3] were observed[9] and correlated with emerging spatial patterns for a very simple "paper, scissors, stone" model where all species are treated symmetrically. In contrast, we investigate a more general model which breaks the symmetry between the species and results in a much richer spatial behavior.

Our analytical results for the inter and intra-species pair correlation functions, supported by computer simulations for long range correlations, show new and interesting behaviors which can be interpreted as strategies of survival for minority agents based on *selection of patterns*. The surviving (stationary) patterns show complex spatial organization which looks similar to what might be expected to emerge if the species were trying to maximize their chances for survival. Although such stationary states are to be expected as means of self-organization of an interacting system, they should not be confused with global equilibrium states. In fact, these states are manifestations of spatial nonequilibrium critical phenomena[10].

We consider a model of a disease spreading through a spatially correlated three-species population evolving in time according to stochastic rules. The population members (agents) occupy the sites of a discrete, two dimensional square lattice with the neighborhood of a given site defined in the von Neumann sense (i.e. each site has four nearest neighbors). Thus, each site of the lattice can be in three different states: empty (*O*), occupied by one healthy agent (*X*) or occupied by one infected agent (*Y*). The time evolution of this system is cyclic (irreversible) and analogous to that occurring in contact processes models[2,10], or in lattice gas models[3] describing position-fixed chemical reactions[8]. The local transition rules are: a given empty site gets occupied by a healthy agent, i.e. $O \to X$, at a rate $\frac{1}{4} a N_X$ where $N_X$ is the number of healthy nearest neighbors of that site. A healthy agent gets infected, i.e. $X \to Y$, at a rate $\frac{1}{4} b N_Y$ where $N_Y$ is the number of infected nearest neighbors, while an infected agent dies, i.e. $Y \to O$, at a rate *d* independent of its neighborhood. This last rule explicitly breaks the symmetry of a basic cyclic system studied previously[9].

Furthermore, depending on a particular application of the model, the three agents can bear different names. For instance, in the context of epidemiology our generic terminology for the trio (healthy, infected, dead) is substituted by the trio (susceptible, infectious, recovered) in direct correspondence with the classical Kermack-McKendrick[7] SIR model. On the other hand, in the mathematical ecology context, the three agents are referred to as (prey, predator, empty) to mimic the classical Lotka-Volterra[6] system.

According to the just described rules of evolution and assuming homogeneity and isotropy of the configurational space, the probability equations for the rate of change of the density $\rho_\alpha$ of species $\alpha = O, X, Y$, are:

$$\begin{aligned}\dot\rho_X &= a P_{OX} - b P_{XY} \\ \dot\rho_Y &= b P_{XY} - d \rho_Y\end{aligned} , \qquad (1)$$

where $P_{\alpha\beta} = P_{\beta\alpha}$ are the joint probability densities for finding a species $\alpha$ at a given site and a species $\beta$ at a nearest neighbor of that site, $\dot\rho_\alpha \equiv \frac{d}{dt}\rho_\alpha$. Since $\sum_\alpha \rho_\alpha = 1$, only two of the species densities are independent. Similarly, since $\sum_\beta P_{\alpha\beta} = \rho_\alpha$, only three out of nine equations for joint probabilities are independent. We choose to work with cross-species probabilities which rates of change equations are:

$$\dot{P}_{OX} = dP_{XY} - \tfrac{1}{4}aP_{OX} + \tfrac{3}{4}\left[a\left(T^O_{OX} - T^O_{XX}\right) - bT^X_{YO}\right]$$
$$\dot{P}_{XY} = -(d + \tfrac{1}{4}b)P_{OX} + \tfrac{3}{4}\left[b\left(T^X_{XY} - T^X_{YY}\right) + aT^O_{XY}\right] ,\qquad(2)$$
$$\dot{P}_{YO} = d(\rho_Y - P_{XY} - 2P_{YO}) + \tfrac{3}{4}\left(bT^X_{YO} - aT^O_{XY}\right)$$

where $T^\beta_{\alpha\gamma}$ denotes the joint probability of finding a species β at a given site and species α and γ on the two nearest neighbors of that site. Subsequently, the time evolution of three-site cluster probabilities can be expressed in terms of four-site cluster probabilities, thus forming the beginning of an infinite hierarchy of coupled equations. These will not be presented here, however, since we shall shortly introduce a truncation scheme cutting the hierarchy at the level of Eqs. (2).

First, let us make further simplifying assumptions. Without much loss in generality we measure the time in units of maximal event rate: $a + b + d = 1$, and we take the rates of infection and recovery to be equal, $a = b = \tfrac{1}{2}(1 - d)$, a choice corresponding to a diagonal cut through the parameter space. Furthermore, we shall concentrate on the study of active stationary states with nonzero population of all three species[3], i.e. the death rate range $0 < d < d_c$. Subsequently, the fixed stationary points of Eqs. (1) are given by:

$$P_{OX} = P_{XY} = \rho_d \rho_Y , \quad \rho_d = \tfrac{2d}{1-d} . \qquad(3)$$

Now, even the mean field approximation, i.e. $P_{\alpha\beta} \approx \rho_\alpha \rho_\beta$, yields three distinct stationary states: two homogeneous absorbing states, one being the all empty lattice for $d = 0$, and the other the all healthy population state for $d > d_c = \tfrac{1}{3}$, and an active endemic state with nonzero population of all three species: $\rho_X = \rho_d, \rho_Y = \rho_O = (1 - \rho_d)/2$, for $0 < d < d_c$.

To improve on the mean field results we define the pair approximation[11,12] via:

$$T^\beta_{\alpha\gamma} \approx P_{\alpha\beta} P_{\beta\gamma} / \rho_\beta . \qquad(4)$$

With the above approximation, the fixed point conditions for pair densities of Eqs. (2), reduce to nonlinear algebraic equations for three unknowns: the cross-species pair probabilities $P_{XY}$ and $P_{YO}$, and the healthy species density $\rho_X$. However, for our purpose, it is more convenient to work with conditional probabilities defined as:

$$C_{\alpha/\beta} \quad P_{\alpha\beta}/\rho_\beta \, , \tag{5}$$

where $C_{\alpha/\beta}$ is the probability of a nearest neighbor to a site in a state $\beta$ to be in a state $\alpha$. In terms of these conditional probabilities, the stationarity conditions of Eqs. (3) are expressed as:

$$C_{X|Y} = \rho_d \, ; \quad C_{O|X} = C_{Y|X} \, ; \quad C_{O|Y} = \rho_d C_{Y|O} C_{X|O}^{-1} \, . \tag{6}$$

The remaining three cross-species probabilities can be calculated from the stationarity requirements of Eqs. (2):

$$\begin{aligned} C_{Y|X}\left[C_{Y|O} + C_{Y|X} + 2C_{X|O} - \tfrac{2}{3}(2\rho_d + 1)\right] &= 0 \\ C_{Y|X}\left[C_{Y|O} - 3C_{Y|X} - \tfrac{2}{3}(2\rho_d - 1)\right] &= 0 \\ C_{X|O}\left[C_{Y|X} - C_{Y|O} - \tfrac{4}{3}(\rho_d - 1)\right] - \tfrac{8}{3}\rho_d C_{Y|O} &= 0 \end{aligned} \tag{7}$$

There are two classes of solutions to Eqs. (7): the absorbing all-$X$ state, and the active state with nonzero average population of all three species. Indeed, since the species densities ratios can be written as:

$$\frac{\rho_\alpha}{\rho_\beta} = \frac{C_{\alpha/\beta}}{C_{\beta/\alpha}} \, ,$$

then with the help of relations Eqs. (6) the species densities are calculated as:

$$\begin{aligned} \rho_X &= \rho_d C_{X|O} \bar{\rho} \\ \rho_Y &= C_{Y|X} C_{X|O} \bar{\rho} \\ \rho_O &= \rho_d C_{Y|X} \bar{\rho} \\ \bar{\rho} &= \left[\rho_d C_{X|O} + (\rho_d + C_{X|O}) C_{Y|X}\right]^{-1} \, . \end{aligned} \tag{8}$$

Clearly, the $C_{Y|X} = 0$ solution of Eqs. (7) yields $\rho_Y = \rho_O = 0$ and $\rho_X = 1$, corresponding to the absorbing state. Moreover, the active state solutions of Eqs. (7) are given by:

$$\begin{aligned} C_{Y|X} &= \tfrac{1}{3}\left\{2 + \rho_d - \sqrt{1 + 4\rho_d + 9\rho_d^2}\right\} \\ C_{Y|O} &= \tfrac{2}{3}(2\rho_d - 1) + 3C_{Y|X} \\ C_{X|O} &= \tfrac{2}{3} - 2C_{Y|X} \end{aligned} \tag{9}$$

Since $C_{Y/X}$ of Eq. (9) vanishes for $\rho_d \ge \sqrt{3/8}$, i.e. for $d \ge d_c \approx 0.234$, the phase transition from the active to the all healthy state takes place earlier than predicted by the mean field approximation where $d_c = \frac{1}{3}$. However, the extinction of the healthy species still occurs only for $d = 0$. The pair approximation results for all three species densities, Eqs. (8) with Eqs. (9), in the active region are plotted in Fig. 1 with solid lines. We note that for low values of the death rate the infected agents density is suppressed, while empty site density is enhanced compared to mean field results.

While the mere existence of the two phase transitions is not surprising in itself (the pair approximation predictions are not qualitatively different from those of the mean field), the spatial correlation functions calculated in the pair approximation, Eqs. (9), offer important clues about the nature of these transitions. In particular, the limits

$$\lim_{d \to 0} C_{X/X} = \frac{1}{3}$$
$$\lim_{d \to d_c} \frac{C_{Y/Y}}{C_{O/O}} = \frac{\frac{1}{4}}{1 - \sqrt{\frac{2}{3}}} \qquad (10)$$

revel strong same-species clustering effects in the limits where the corresponding agents are in minority. Moreover, the non-zero limits

$$\lim_{d \to d_c} \frac{C_{Y/O}}{C_{O/Y}} = \frac{\sqrt{\frac{2}{3}} - \frac{2}{3}}{\frac{3}{4}\left(1 - \sqrt{\frac{2}{3}}\right)} \qquad (11)$$

are indicative of "cooperative" correlations among agents which are simultaneously on the verge of extinction near the upper phase transition. Thus, the pair approximation results already suggest a "defensive" spatial organization of the minority species near their respective extinction limits. Note that as expected, the mean field limits of Eqs. (10) and (11) are all equal to zero.

To emphasize the spatial nature of the inter and intra-species correlations it is illustrative to define the correlation strength as

$$I_{\alpha/\beta} \equiv \rho_\alpha - C_{\alpha/\beta} , \qquad (12)$$

which measures the deviation of the nearest neighbor pair correlations from that expected for a random distribution. The pair approximation results of Eqs. (8) and (9) for these correlation strengths (interactions) are displayed in Fig. 2 for the central site of the healthy type and in Fig. 3, for the central site in the infected state. The remaining three correlation functions with the central site being empty are not shown since these are nearly identical to those of Fig. 3. With the above definition, Eq. (12), the positive value of $I_{\alpha/\beta}$ signifies a repulsive interaction, while negative values correspond to attraction.

Consequently, all three same-specie interactions are attractive in the entire range of the active state, while the cross-specie interactions are repulsive except for the infected-dead correlations, which change sign for $d \approx 0.1$ (middle of the range of the active state). This change of sign is to be expected if one interprets the mutual inter-specie correlations as emergent "strategies for survival" of the minority agents. At low death rate values the healthy agents are in minority. The highest "survivability" patterns for healthy species are then those that tend to minimize the length of the contact boundary with the "invading" infected species: healthy agents cluster together attracting each other and repelling the rest. The infected agents tend to surround healthy clusters, repelling the dead in the process. For high death rates, the infected and dead agents both are in minority and thus tend to "cooperate" in their spatial organization, acting as a single species (attractive infected-dead interaction) in defense against healthy invaders. The resultant clustering of the minority agents in this range of the death rate is expected to be weaker, however, since the "cooperating" species have distinct "goals". To increase in numbers, the infected agents need to maximize contacts with the healthy majority while the dead need to minimize their contacts with the healthy agents. The ensuing compromise weakens the clustering strategy.

We test the accuracy of the pair approximation, Eq. (4), by performing computer simulations on a $100 \times 100$ lattice. Both periodic and fixed-end boundary conditions were employed and our results are not particularly sensitive to the choice of boundary conditions. We take one time step to correspond to $10^4$ updates of individual sites, i.e. approximately one update per site. Spatial correlations are found by averaging over time. The system was started in a variety of initial configurations and all averaging takes place after the first 500 time steps (minimum) so as to exclude initial fluctuations. The total number of time steps included in the average ranged from 4,000 to 12,000. In fact, the pair correlation functions are found to be reasonably accurate in just 1,000 steps. The long range correlations, to be discussed later, become significantly less noisy when averaging over 12,000 steps, although the basic trends are well established after just 4,000 steps.

The results of the above simulations are displayed in Figs. 1, 2, and 3, in direct comparison with the pair approximation, showing excellent qualitative agreement. Quantitatively, however, the pair approximation consistently underestimates the strength of correlations close to both lower and upper phase transitions, suggesting the emergence of long range correlations. Furthermore, the critical values of the death rate shift according to the simulations: $d_c \approx 0.18$, while the extinction transition occurs at a non-zero value $0 < d \approx 0.025$. However, the latter critical value shows sensitive dependence on the lattice size, decreasing with increasing size[3]. More studies are needed to determine whether in the limit of infinite system size the extinction of the healthy species happens only at vanishing death rate as in both pair and mean field approximations.

To test the conjecture of emergent long range interactions beyond the pair approximation, we have performed further simulations. The long range spatial correlations between species are again measured with respect to a random distribution and are defined as:

$$I_{\alpha/\beta}(n) = \rho_\alpha - C_{\alpha/\beta}(n) , \qquad (13)$$

where $C_{\alpha/\beta}(n)$ denotes conditional probability of a site distance $n$ away from a chosen site to be in a state α, subject to the chosen site being in a state β. Thus, the nearest neighbor pair correlation $I_{\alpha/\beta} \equiv I_{\alpha/\beta}(1)$. The numerical simulation results for the three most interesting correlation functions are displayed in Fig. 4 for four characteristic values of the death rate.

In general, we see that the correlation strength extends to a substantial number of sites for death rates close to the lower phase transition, i.e. $d \approx 0$. As $d$ is increased, the range of interactions is significantly reduced. Specifically, the same-species interaction (*X/X*) is attractive at low *d* values up to a distance of 20 sites. Similarly, for low *d*, the dead-infected (*O/Y*) interaction is strongly long range repulsive. For large death rates (close to upper phase transition), the *O/Y* interaction becomes weakly attractive, consistent with the predictions of the pair approximation calculations for the cooperating minority species.

Of particular interest in Fig. 4, is the infected-dead (*Y/X*) correlation, which clearly shows attractive cross-species interactions 5-10 sites away from the central site, despite the fact that the nearest neighbor correlation is repulsive. This attractive correlation at longer distances can be quite significant. For $d = 0.025$, the strength of this attractive interaction at $n = 6$ is about 50% of the repulsive nearest neighbor correlation. Therefore, the combined effect of all the inter and intra-species correlations is to preferentially select the highest survivability patterns. For $d \approx 0$, these patterns typically include clusters of species *X* (healthy or prey) which are surrounded by a region with a surplus of species *Y* (infected or predators), while the O species (dead or empty) are pushed farther away from the central clusters. We have observed the prevalence of such patterns in real time computer simulations.

This global picture of pattern selection is reminiscent of defensive strategies typically associated with learned or instinctual behavior of minority (or weaker) species. For instance, in the context of the predator-prey model, prey assemble in clusters for protection while predators surround their prey and avoid empty space. It is interesting to notice that these complex strategies emerge in our model, despite the model being built on simple contact interactions only.

The authors acknowledge support from the Global Change and Environmental Quality Program of the University of Colorado and ARO Grant # DAAH04-94-G-0253.

# FIGURES

Fig. 1. Active state specie densities $\rho_\alpha$, $\alpha = X, Y, O$, according to the pair approximation (solid lines), and corresponding numerical simulation results (open symbols).

Fig. 2. Pair correlation function of the healthy species $I_{\alpha/X}$, $\alpha = X, Y, O$, according to the pair approximation (solid lines), and corresponding numerical simulation results (open symbols).

Fig. 3. Pair correlation function of the infected species $I_{\alpha/Y}$, $\alpha = X, Y, O$, according to the pair approximation (solid lines), and corresponding numerical simulation results (open symbols).

Fig. 4. Numerical simulation results for the interaction strength $I_{\alpha/\beta}(n)$ for some $\alpha, \beta = X, Y, O$ as a function of site index $n$. Corresponding values of the death rate $d$ are indicated in the panels.

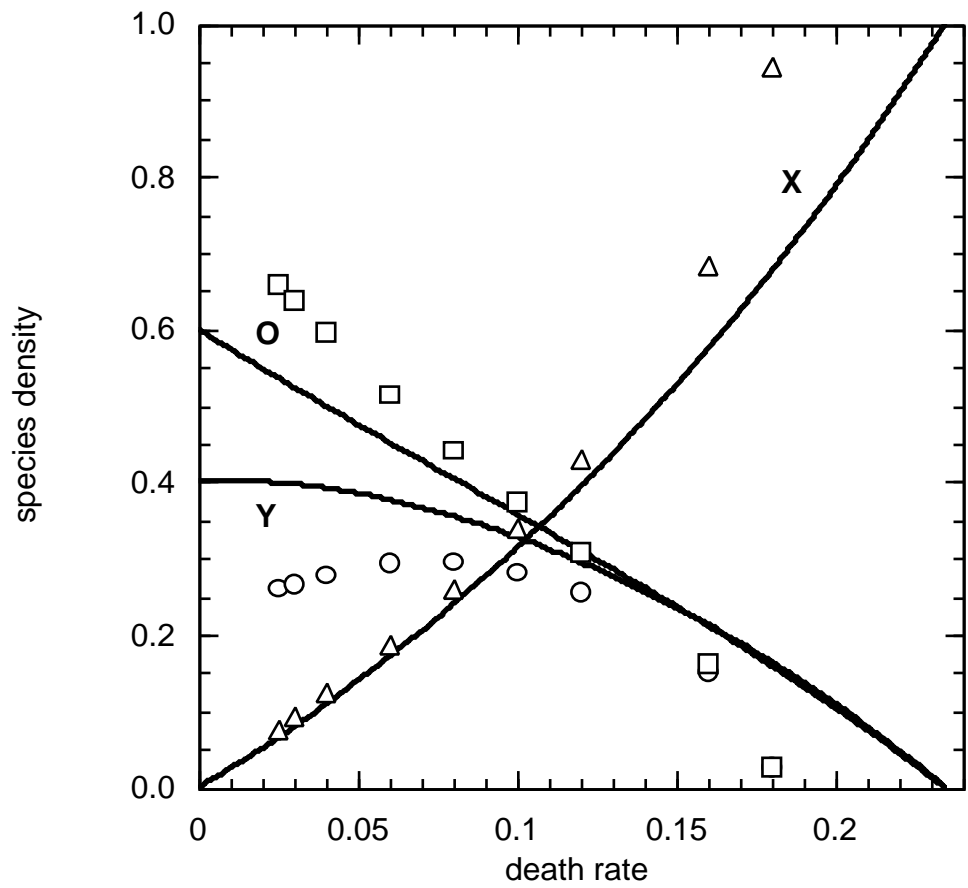

Fig. 1

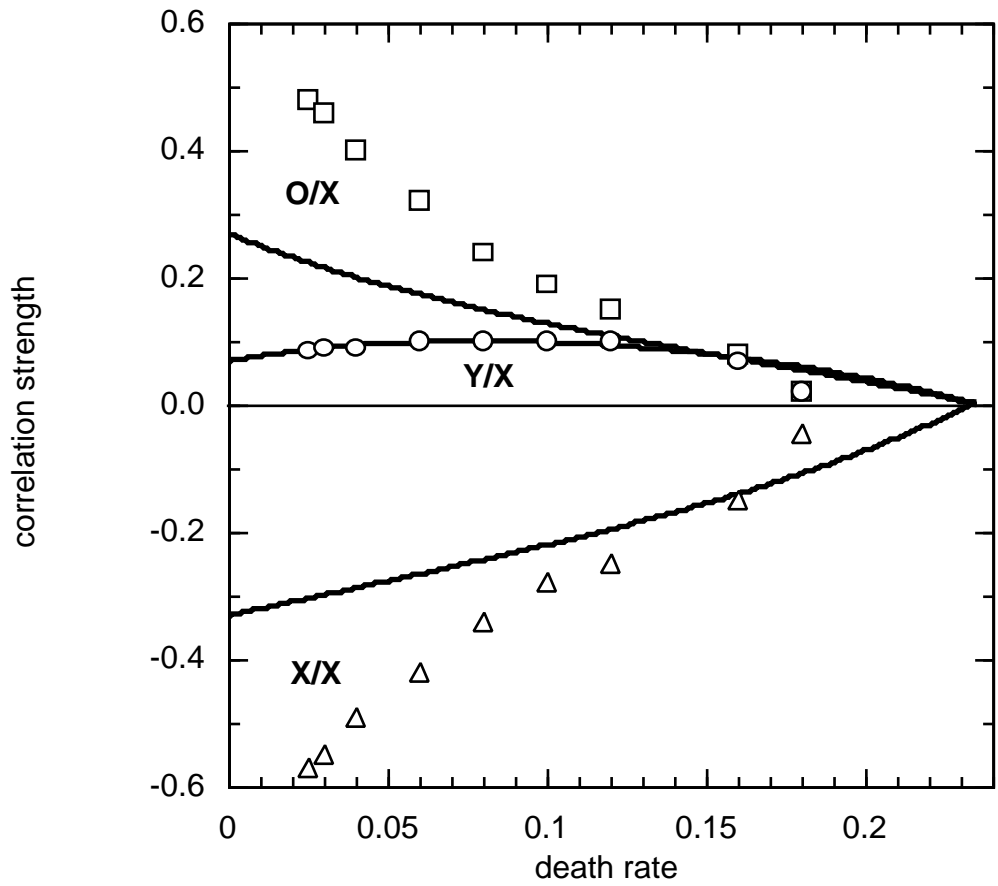

Fig. 2

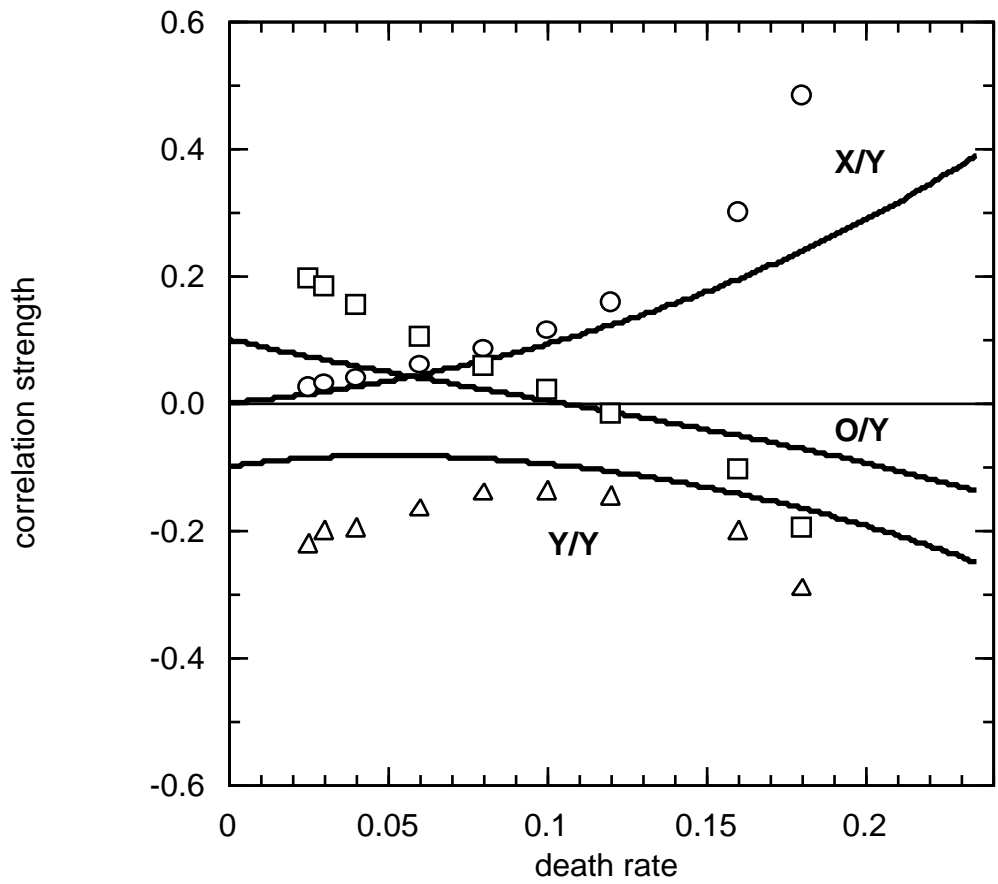

Fig. 3

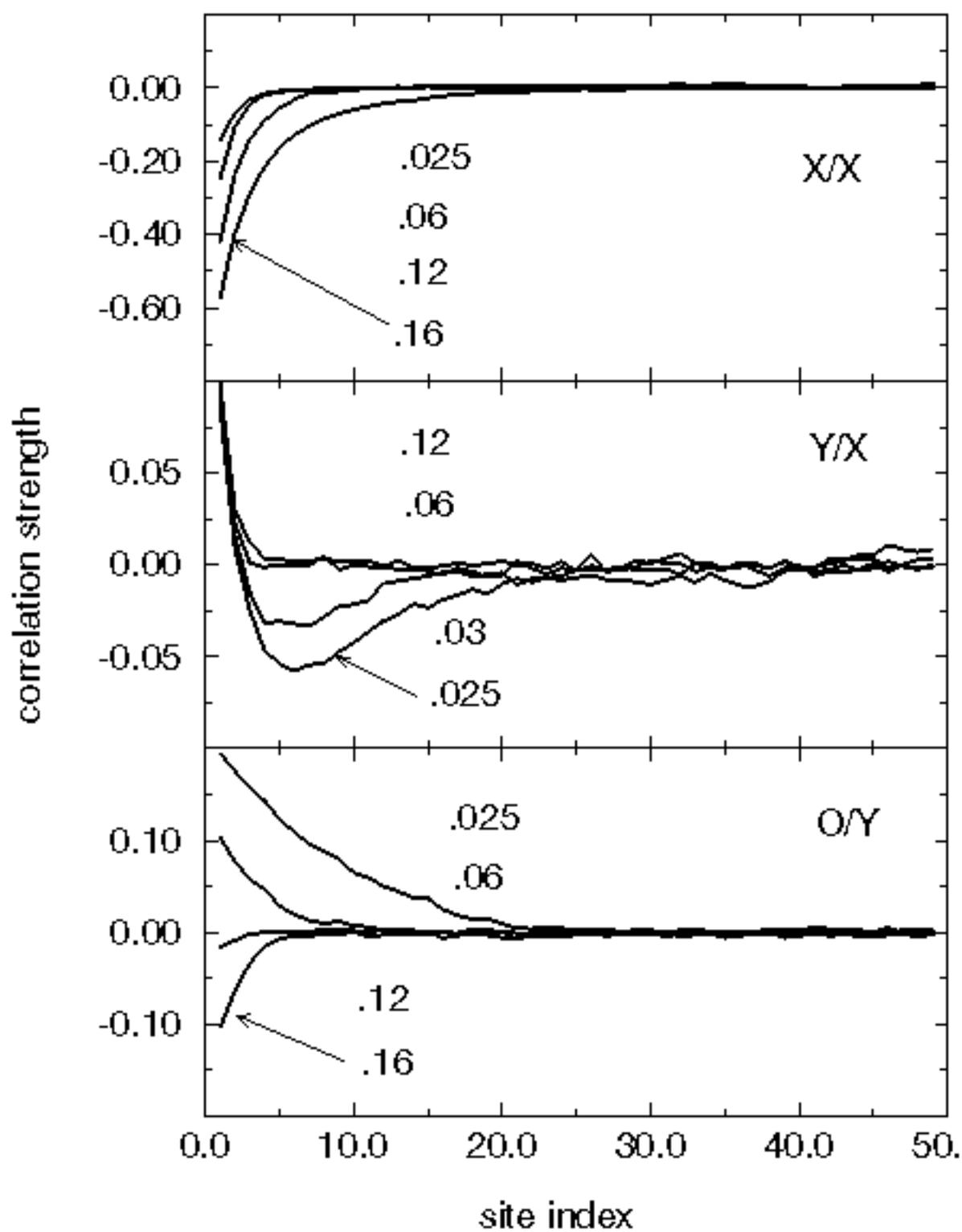

Fig. 4